# COMPREHENSIVE ANALYSIS ON DETERMINANTS OF BANK PROFITABILITY IN BANGLADESH

Md Saimum Hossain[1]
Faruque Ahamed[2]

**Abstract**

The study investigates the relationship between bank profitability and a comprehensive list of bank-specific, industry-specific and macroeconomic variables using unique panel data from 23 Bangladeshi banks with large market shares from 2005 to 2019 employing the Pooled Ordinary Least Square (POLS) Method for regression estimation. The random Effect model has been used to check for robustness. Three variables, namely, Return on Asset (ROA), Return on Equity (ROE), and Net Interest Margin (NIM), have been used as profitability proxies. Non-interest income, capital ratio, and GDP growth have been found to have a significant relationship with ROA. In addition to non-interest income, market share, bank size, and real exchange rates are significant explaining variables if profitability is measured as NIM. The only significant determinant of profitability measured by ROE is market share. The primary contribution of this study to the existing knowledge base is an extensive empirical analysis by covering the entire gamut of independent variables (bank-specific, industry-related, and macroeconomic) to explain the profitability of the banks in Bangladesh. It also covers an extensive and recent data set. Banking sector stakeholders may find great value from the outputs of this paper: Regulators and policymakers may find this useful in undertaking analyses in setting policy rates, banking industry stability, and impact assessment of critical policy measures before and after the enactment, etc. Investors and the bank management are to use the findings of this paper in analyzing the real drivers of profitability of the banks they're contemplating to invest and managing on a day-to-day basis.

## I.   INTRODUCTION

From a global perspective, banks' profitability and drivers have been a topic of curiosity among academicians and practitioners. Several studies have been carried out to pinpoint the factors capable of explaining the profitability of banks and the exact nature of the relationship between these two sets of variables. The current study is an addition to the long list of studies already available on this topic. However, it focuses on doing comprehensive research on the banking sector of Bangladesh.

---

[1] Assistant Professor, Department of Finance, FBS, University of Dhaka.
[2] Department of Economics, Northern Illinois University.

This study finds it relevant because Bangladesh has a primarily bank-dominated financial system since its birth in 1971.

The Bangladeshi banking sector has seen many changes, from the nationalization of all banks to financial liberalization. Bangladesh nationalized all the banks right after the liberation war of 1971. However, Private Commercial Banks (PCBs) have been being patronized since 1981. Table 1 shows a matrix of the different types and number of banks in the country:

**Table 1: The Structure of the Bangladesh Banking Sector[3]**

| Total Scheduled Banks (61) | | |
|---|---|---|
| *State-owned (06)* | *Privately-owned (43)* | |
| Scheduled banks (06) | Locally owned (40) | NRB banks (03) |
| Specialized banks (03) | Conventional banks (33) | |
| | Shariah-based Islamic banks (10) | |
| | Foreign-owned (09) | |

Apart from the above, five non-scheduled banks and 34 non-bank financial institutions operate in the country.[4]

Studies on determinants of bank profitability can be categorized across two types of variables: the development stage of the economy and the type of the variable used to explain the profitability measure. Table 2 below shows the common types found in earlier literature:

**Table 2: Common Study Types on Bank Profitability and Its Determinants**

| Development-stage of the economy | Type of variable used | Geography-based |
|---|---|---|
| Developed/High-income | Bank-specific/internal variables | Single country |
| Developing/Middle-income | Industry-specific/external variables | Multi-country |
| Emerging/Lower-middle-income | Macroeconomic/environmental variables | Multi-continent |

---

[3] https://www.bb.org.bd/fnansys/bankfi.php
[4] ibid

There's a dearth of studies on the profitability determinants of the Bangladesh banking sector. The ones available either cover a relatively older data set or test only one category of independent variables against the profitability measures. The primary contribution of this study to the existing knowledge base is an extensive empirical analysis by covering the entire gamut of independent variables (bank-specific, industry-related, and macroeconomic) to explain profitability of the banks in Bangladesh. It also covers a relatively wide (15 years) and recent data set (till 2019). This combination of extensive data and a comprehensive selection of bank-related, industry-specific and macroeconomic factors enable us to accomplish this objective. The output from this paper is a valuable tool for related stakeholders. Regulators and policymakers may find this useful in undertaking analyses in setting policy rates, banking industry stability, and impact assessment of critical policy measures before and after the enactment etc. Investors and the bank management are to use the findings of this paper in analyzing the real sources/drivers of profitability of the banks they're contemplating to invest and managing on a day-to-day basis.

This paper is organized as follows: Section II covers a review of the relevant literature, section III presents the methodology, including the econometric model, data, and empirical model description, section IV discusses the results, and section V concludes the paper and suggests scope for further research.

## II. LITERATURE REVIEW

We can classify the previous literature on bank profitability determinants from several angles: some researchers have studied only bank-specific or internal variables. Some others have studied external (industry-related and macroeconomic) variables and their effect on bank profitability. Many studies cover single-country data, whereas studies done in a multi-country or multi-continent setting are also common. Lastly, researchers focused on undertaking studies based on the life-cycle stage of

economies – developed, developing, emerging, etc. We present a review of the previous literature based on geographic classification:

*1. Single-country studies*

Studies looking at both bank-specific and macroeconomic factors are widespread for different economies in the world. For bank-specific factors which have a strong influence on the profitability of banks, studies were conducted by Bhatia et al. (2012); Sufian and Noor (2012) in India; Liu and Wilson (2010) in Japan; Shoaib et al. (2015) in Pakistan; Sufian and Chong (2008) in the Philippines; Macit (2012); Alper and Anbar (2011); Alp et al. (2010) in Turkey; Kosmidou et al. (2005); Sufian (2011) in Korea, Saeed (2014) in the United Kingdom, etc.

Sufian (2011) used 251 bank information of Korea from 1992-2003 and found that liquidity had negative and noninterest income has a positive relationship with profitability. Goddard, Molyneux, and Wilson (2004) conclude that banks with higher liquidity witness lower profits. Al-Jarrah et al. (2010) conducted a study using the cointegration and error correction models to identify the determinants of profitability on all Jordanian banks over 2000- 2006. According to the study, loans and advances outstanding to total assets ratio, noninterest or operating expenditures ratio, the capital arrangement, and the deposit to asset ratio are important internal determinants of profitability.

Macit (2012) conducted a study using quarterly unconsolidated balance sheets of participating banks that operated between 2005 and 2010 in Turkey. The study found that the equity to total asset ratio has a positive impact on profitability. In contrast, the ratio of nonperforming loans to total outstanding loans and advances has a negative relationship. Gul et al. (2011) used the pooled Ordinary Least Square (POLS) method to identify the relationship between bank-specific and macroeconomic characteristics over bank profitability by using data of top 15 Pakistani commercial banks over 2005-2009. They identified that assets, loans, equity, and deposits positively impact all three profitability indicators, i.e., ROA, ROE, and NIM.

Shoaib et al. (2015) conducted a study through the POLS regression model by using the panel data of all scheduled banks of Pakistan from 2006-2013. The empirical results show that banks' profitability is adversely affected by liquidity, nonperforming loans, and administrative expenses and positively affected by capital adequacy. An increase in operating expenses causes the profitability of Turkish banks to fall, commented Alp et al. (2010). They also identified that there does not exist any statistically significant relationship between total loans and receivables to total assets ratio with the indicators of profitability.

Growe et al. (2014) conducted a study during 1994-2011 over U.S. regional banks using the Generalized Method of Moments (GMM) estimator technique. They found that the level of nonperforming assets is negatively related to all measures of profitability. Acaravci and Çalim (2013) explained that in the case of private commercial banks, the volume of deposits has an insignificant impact on profitability, and higher nonperforming loans reduce profitability by a large extent. In contrast, capital adequacy has a significant and positive impact on profitability. According to Hassan and Bashir (2003), bank profitability measures respond positively to the increases in capital.

Kosmidou et al. (2005) studied U.K.-owned commercial banks during 1995-2002 to identify bank-specific characteristics, macroeconomic conditions, and financial market structure on banks' profits and found that capital strength and efficiency in expenses management positively and leading influence on their performance. Kosmidou (2006); Pasiouras et al. (2006) reveal an adverse effect of liquidity on bank profitability. Vieira (2010) found a weak short-run positive relationship between ROA and liquidity. According to Lee and Hsieh (2013); Menicucci and Paolucci (2016), a high volume of deposits leads to higher profits. Similar results were found by Saeed (2014) in his study. However, Demirguç-Kunt and Huizinga (1998) found a mixed relationship between deposit and profitability.

Saeed (2014) investigated the impact variables of profitability on 73 U.K. commercial banks from 2006 to 2012 and concluded that capital ratio, loan outstanding, the volume of deposit deposits, amount of liquidity, and interest rate positively impact ROA ROE. Sufian and Chong (2008) examine the performance determinants of banks in the Philippines during the period 1990–2005. The study suggests that operating expense is negatively related to ROA and ROE while the capital and Non-interest income positively impact profitability.

Bhatia et al. (2012) tried to examine the private sector banks' profitability determinants from 2006-07 to 2009-10. Backward Stepwise Regression Analysis has been conducted on 23 banks to identify the relationship of these determinants and banks' performance. The study reveals that loan and advances outstanding to deposit ratio, Capital adequacy ratio, and non-interest income directly impact Return on Assets. In another study in the Indian banking sector from 2000 to 2008, Sufian and Noor (2012) liquidity and operating expenses significantly impacted profitability.

Batten and Xuan (2019) conducted a study on Vietnam using the panel data method that suggested a substantial impact on profitability from variables like bank size, risk, expense, productivity, capital adequacy, etc. In contrast, industry-related features and macroeconomic variables negatively affect the profitability measures of a bank. Besides, the causality direction is not consistent across profitability measuring proxies.

Rani and Zergaw (2017) conducted their study on Ethiopian banks using multiple regression models to analyze the bank-specific and industry and macroeconomic specific determinants of profitability. The study showed a negative impact of internal and industry-related variables on profitability. In contrast, macroeconomic determinants showed a positive but somewhat insignificant relationship with the net profit margin of the Ethiopian banks.

Bolarinwa et al. (2018) conducted a study on Nigeria using the system generalized method of moments, which showed that cost-efficiency works as a strong determinant in attaining profitability in developing countries. Hasanov et al. (2018) conducted their study in Azerbaijan, which carries an oil-dependent economy implementing the Generalized Method of Moments that indicated internal and external variables like bank size, asset, and liability, oil price, inflation rate, economic cycle, etc. have a positive relationship with profitability. On the other hand, deflation of the exchange date, amount of deposit, and risk regarding the liquidity can negatively affect profitability measures.

Topak and Talu (2017), based their study on Turkey implementing the balanced panel data from 2005 to 2015, find a significant and positive impact of bank-based variables like net interest margin, commissions, etc., on profitability in return on assets and equity. On the other hand, the ratio of NPL and other operating expenses, capital adequacy have a negative relationship with profitability.

Belke and Unal (2017) conducted their study on 23 deposit banks in Turkey using the panel regression method. According to the study, bank size, capital, inflation rate, economic growth, market concentration, exchange and policy rate, etc., have a significant impact on bank profitability. However, the impact and influence differ in terms of listed and non-listed banks. Hasan et al. (2020) demonstrated the effect of bank profitability in terms of two variables: return on asset and return on equity following the Model Panel Data methodology in the context of Indonesia. For the return of equity, variables like net interest margin, capital adequacy ratio, loan to deposit Ratio etc., be significant.

Ali and Puah (2018) conducted a panel regression analysis of 24 Pakistani commercial banks for the 2007-2015 periods and found a statistically substantial impact of bank size, credit and funding risk on profitability. Liquidity risk had no significant statistical impact. Another study on bank profitability in Sri Lanka was approached by Kawshala and Panditharathna (2017), implementing the panel data method on 12 Sri Lankan domestic, commercial banks. The study reveals that variables such as capital ratio, deposit ratio, etc., have a significant and positive relationship with bank profitability and liquidity negatively associated with profitability.

*2. Multi-country studies*

Mauricio et al. (2014) found a positive relationship between capital adequacy and profitability by using the panel data of 78 commercial banks from Argentina, Brazil, Chile, Colombia, Mexico, Paraguay, Peru, and Venezuela over the period from 1995 to 2010.

Empirical evidence by Demirguc-Kunt and Huizinga (1999) suggests that banks that preserve higher equity levels compare to their assets tend to perform better. Goaied (2008); Pasiouras and Kosmidou (2006); Dietrich and Wanzenried (2009); Obamuyi (2013); Garcia-Herrero et al. (2009); Menicucci, and Paolucci (2016) found that higher equity ratio on total assets can be an essential factor on the profitability of banks in Europe.

Sahyouni and Wang (2018) conducted their study using the panel data fixed effect technique on 11 developed and emerging countries for the 2011-2015 period. They concluded that management, capital ratio, and bank size indicate a positive relationship with profitability, whereas banks that generate higher liquidity are likely to achieve lower profitability.

Boateng (2018) conducted a comparative study on 20 India and Ghana-based banks (10 banks from each country) using the multiple regression method. According to the study, macroeconomic and bank-specific variables like credit risk, net interest margin, liquidity, capital adequacy ratio, bank size, etc., had a remarkable impact on the profitability measure (return on asset) of Indian and Ghanaian banks. However, bank size and cost to income ratio had a significant effect on Ghana's profitability rate and comparatively insignificant influence in terms of India.

Özsarı, et al. (2018) conducted their research on 13 post-Soviet countries using the Generalized Method of Moments and panel regression. They found a positive relationship of economic growth and non-interest bank loan with profitability and a negative association of loan-to-GDP with profitability.

*3. Bangladesh studies*

Islam and Rana (2017) conducted their study on 15 selective private banks of Bangladesh using panel data focusing on internal variables affecting bank profitability measures. They find a strong negative impact of operating expenses and nonperforming loans (NPL) on bank profitability.

Mahmud et al. (2016) conducted a study covering bank-related data for the 2003-2013 period and found size, operating expenses ratio, and gearing ratio to negatively affect profitability.

Using the Generalized Method of Moments (GMM) for a 2006-2013 panel data set, Rahman et al. (2015) conclude that capital and loan intensity have positive, and cost efficiency and off-balance sheet activities have a negative relationship with bank profitability.

Taking a 2012-2016 dataset of the top 15 private commercial banks in Bangladesh by asset size, Hossain and Ahamed (2015) find that bank earnings, asset quality, management efficiency, capital strength, size, and asset structure have a significant impact on bank profitability. In another study, Ahamed (2021) found a positive correlation between liquidity and profitability using the annual data for 2005-2018.

Sufian and Habibullah (2009) took data from 1997 to 2004 and analyzed it using the Lease Squares and Fixed Effect model. They find a negative correlation between profitability and a bank-specific variable (bank size) and a macroeconomic variable (inflation).

Dey (2014); Sufian & Habibullah (2009); Abdullah et al. (2014) conducted studies with a similar focus on a specific cluster of variables in explaining bank profitability.

## METHODOLOGY

*Data*

The study investigates the relationship between bank profitability and bank-specific, industry-specific, and macroeconomic variables. The data is collected from the annual financial statements of the Banks listed in the Stock exchange for the period of 2005-2019, which are available in the Bangladesh Securities and Exchange Commission library. The macroeconomic information is retrieved from the Bangladesh Bureau of Statistics, Bangladesh Bank, IMF Financial Statistics, and World Bank database. Data of 23 banks with large market shares for the years 2005 to 2019 has been used, giving 245 bank-year observations.

*Econometric Model*

Panel data has been used to measure the cross-section units' variability and dynamic change over time. A pooling analysis allows obtaining more consistent estimates of the parameters where the association between the variables is stable through cross-section units. Pooled Ordinary Least Square (POLS) Method displays the general quality of minimized bias and variance, which is considered the most consistent regression estimation. Demirguc-Kunt and Huizinga (1999); Short (1979); Bourke (1989); Molyneux and Thorton (1992); Menicucci and Paolucci (2015) use the simple linear equation model to analyze the relationship with profitability.

To identify the relationship between the profitability of bank and the bank-specific, industry-specific and macroeconomic variables we estimate the following linear regression model:

$$Y_{ij} = \alpha + \beta_1 NII_{ij} + \beta_2 DPST_{ij} + \beta_3 OPEX_{ij} + \beta_4 CAPR_{ij} + \beta_5 LTAR_{ij} + \beta_6 SIZE_{ij} + \beta_7 MKT_{ij} + \beta_8 INF_{ij}$$

$$+ \beta_9 GDP_{ij} + \beta_{10} EXH_{ij} + \epsilon$$

In this equation *i* refers to a specific bank, *j* refers to a year, $Y_{ij}$ refers to bank profitability and is the observation of bank *i* in a particular year *j*. and $\epsilon$ is a normally distributed random variable disturbance term or error term with zero variance.

*Dependent variable(s)*

In the literature, three measures of profitability, such as Return on Assets (ROA), Return on Equity (ROE), and Net Interest Margin (NIM), are used and expressed as a function of the internal and external determinants.

Return on Assets (ROA) shows the profit of the company over its assets. It measures the efficiency of utilizing assets to generate income for a company. ROA is a better measure of the ability of the firm to generate returns on its portfolio of assets. ROA is used to identify the operational performance, competence, and efficiency of a bank. ROA is used by many researchers in previous literatures.

Return on Equity (ROE) measures the amount of income a company generates against its equity. It explains how effectively managing a company is using the shareholders' equity capital to earn profit. ROE does not account for financial leverage, so the ratio tends to be higher than the ROA. ROE measures how successful a company uses its investment funds to cause earnings growth.

The net Interest Margin (NIM) variable is calculated by dividing the net interest income by the total assets. The net interest income is found by subtracting the total interest expense from the total interest income. NIM is a good measure of profitability as it shows the interest profit earned by the bank by using funds of the depositors and shareholders.

## Table 3: Variables

| Variables | | Measure | Proxy | Hypothesized Relationship |
|---|---|---|---|---|
| *Dependent Variable* | | | | |
| ROA | ROA | Net Profit/Total Asset | Profitability | N/A |
| ROE | ROE | Net Profit/Total Equity | | N/A |
| NIM | NIM | Net Interest Margin/Total Asset | | N/A |
| | | | | |
| *Independent Variables* | | | | |
| Bank-specific Variables | | | | |
| NII Ratio | NII | Non-interest Income/Total Asset | Earnings | + |
| DPST Ratio | DPST | Total Deposit/Total Asset | Asset Structure | + |
| OPEX Ratio | OPEX | Operating Expense/Total Asset | Management Efficiency | +/- |
| CAP Ratio | CAPR | Total Equity/Total Asset | Capital Strength | + |
| LTA Ratio | LTAR | Loans and advance/ Total Asset | Liquidity | +/- |
| | | | | |
| Industry-specific Variables | | | | |
| SIZE | SIZE | Natural Logarithm of Total Asset | Industry Impact | +/- |
| MKTSHARE | MKT | Bank Asset/Total Banking Asset | Market Share | +/- |
| | | | | |
| Macroeconomic Variables | | | | |
| INF Rate | INF | Inflation Rate | Inflation | +/- |
| GDP Growth | GDP | GDP growth rate | Economics Growth | +/- |
| REX Rate | EXH | Exchange Rate | Real Exchange Rate | +/- |

*Independent variables*

Internal determinants involve the factors influenced by the bank management's decisions, efficiency, policy, and objectives. The external determinants can be comprised of both industry and macroeconomic variables that display the banks' economic, legal, and competitive environments. The independent variables fall into three categories of bank-specific variables, industry-specific variables, and macroeconomic variables. All these variables have an independent effect on the profitability of the bank.

1. *Bank specific variables:*

The bank-specific variables include earnings, asset structure, asset quality, management efficiency, liquidity, and capital strength. Non-interest income has been used to measure earnings. Operating expenditure, total deposit, nonperforming loans, outstanding loans, and shareholders' equity are used to measure the management efficiency, asset structure, asset quality, liquidity analysis, and capital strength, respectively. It is noteworthy that the bank-specific variables are scaled using comprehensive variables like Total Asset or Total Loans to create comparability of data for the sample banks.

Non-interest income is the proxy variable of earnings. Exchange and brokerage commission, fees, investment income, foreign exchange profit, service charge, dividend income, gain from the asset sale, etc., are considered the source of non-interest income. Deposit is the primary source of bank funding, so it is directly correlated with bank profitability. The deposits are used as a proxy of the bank asset structure. It shows the diversification of the assets of the banking business. The more the deposit amount, the higher the opportunity to earn profit by disbursing loans and advances. Nonperforming loans are considered loans and advances which do not generate any income for the bank. Bank must keep provisions from profit against the nonperforming loans. The loan loss provision reduces the distributable profit of the bank. The loan loss provisions reduce the liquidity and affect the disbursement ability of new loans and investments. Operating expense or the Non-interest expense is the measurement of operational efficiency of the management. Equity is used as a proxy of total capital and defines the general safety and soundness of the financial institutions. Higher equity indicates the ability of the bank to absorb losses and handle significant threat and vulnerabilities arising from the business operation. Large, capitalized banks are able to absorb shocks at different levels from various risk factors and perform well in the long run. Loans and Advances are the primary sources of earnings for the banks. Generally Loans and Advances are less liquid than other asset components hence higher loans and advance to asset ratio implies less liquidity.

*2. Industry specific variables:*

Bank size variable is one of the crucial factors that impact the profitability. Several previous studies in empirical research found that size is a determinant of bank profitability. Large banks have strong capital and asset base which allows them to disburse more loans and invest in various securities. Greater market share increases efficiency, generate fund at lower costs and poses strong market power. However, excessively large size of a bank can lead to greater inefficiency and rise agency cost. Market Share ratio can be calculated by dividing the individual bank assets with the total banking asset. This variable identifies the effect of competition in the banking industry.

*3. Macroeconomic variables:*

Macroeconomic factors are considered as the signaling points of economic growth of the country. Macroeconomic variables influence by the government policy, regulations and other overseas factors. These elements are outside the control of the bank and can impact the whole industry to a large extent.

Gross Domestic Product (GDP) growth rate defines the how fast the economy is growing. GDP growth is the increase in the inflation-adjusted market value of the goods and services over the period. Positive GDP growth rate express the economic expansion and progress. Economic growth creates investment opportunities and allows the bank to expand the banking services. Inflation is considered as the sustained increase of price level over a period. Inflation is measured by the change in consumer price index. Inflation can increase the price of the factors of production and therefore raise the cost of business and reduces the profit. The real effective exchange rate is considered as the weighted average of a country's currency relative to the basket of other major currencies such as US Dollars, Euro, Pound and Yen. The real effective exchange rate adjusted for the effects of inflation.

*Robustness Check*

The basic estimation strategy is to pool the observation across banks and apply the regression analysis on the pooled sample. The study uses the least squares method of the random effects (RE) model where the standard errors are calculated by using White's 1980 transformation to control for cross sectional heteroskedasticity. The random effects model has been chosen over fixed effects model by using the Hausman test.

### III.    EMPIRICAL RESULTS AND DISCUSSION

In the study, a total of three models have been developed considering the endogenous variable ROA, ROE and NIM as proxy for profitability. The output from both the random effect model and pooled ordinary least square method depicts consistency, proving the robustness of the dataset. The strong R-squared and adjusted R-squared suggests that all three models explained most of the variation of bank specific, industry specific and macroeconomic variables.

*Model One: Profitability Proxy ROA*

In model one, POLS finds that non-interest income, capital ratio and bank size have positive relationship whereas deposit size, operating expense, loan outstanding, market share, inflation, GDP growth and exchange rate has negative relationship with ROA. However, among the explanatory variables non-interest income, capital ratio and GDP growth has significant relationship. In case of RE model non-interest income, operating expense, capital, market share and bank size positively related with profitability whereas deposit size, loan outstanding, market share, inflation, GDP growth and exchange rate depicts negative relationship. RE models also finds non-interest income, capital ratio, market share, GDP growth and exchange rate to be significant for return on assets.

**Table 4: Model One Estimation Results**

| Variable | Pooled Ordinary Least Square | Random Effect Model |
| --- | --- | --- |
| NII | 0.1703** (0.0547) | 0.1770** (0.0460) |
| DPST | -0.0047 (0.0111) | -0.0069 (0.0092) |
| OPEX | -0.0017 (0.0591) | 0.0377 (0.0506) |
| CAPR | 0.0423** (0.0103) | 0.05792** (0.0092) |
| LTAR | -0.0011 (0.0018) | -0.0011 (0.0017) |
| MKT | -0.07224 (0.084) | -0.0931* (0.0461) |
| SIZE | 0.0002 (0.001) | 0.0003 (0.0010) |
| INF | -0.015 (0.0366) | -0.0226 (0.0367) |
| GDP | -0.2163* (0.0878) | -0.1893* (0.0877) |
| EXH | -0.022 (0.0113) | -0.0253* (0.0113) |

**1% Significance
*5% Significance

**Table 5: Correlation Matrix of Explanatory Variables**

|      | ROA     | NII     | DPST    | OPEX    | CAPR    | LTAR    | MKT     | SIZE    | INF     | GDP    | EXH   |
|------|---------|---------|---------|---------|---------|---------|---------|---------|---------|--------|-------|
| ROA  | 1.000   |         |         |         |         |         |         |         |         |        |       |
| NII  | 0.2541  | 1.000   |         |         |         |         |         |         |         |        |       |
| DPST | -0.0813 | -0.1068 | 1.000   |         |         |         |         |         |         |        |       |
| OPEX | 0.0986  | 0.1775  | -0.0689 | 1.000   |         |         |         |         |         |        |       |
| CAPR | 0.3387  | 0.0783  | -0.2044 | 0.0401  | 1.000   |         |         |         |         |        |       |
| LTAR | -0.0543 | -0.0377 | 0.0660  | -0.102  | -0.099  | 1.000   |         |         |         |        |       |
| MKT  | -0.1203 | -0.0283 | 0.0609  | 0.0712  | -0.0269 | -0.116  | 1.000   |         |         |        |       |
| SIZE | -0.096  | -0.0003 | -0.1464 | -0.0026 | 0.1371  | -0.1318 | 0.3773  | 1.000   |         |        |       |
| INF  | 0.1104  | 0.0820  | 0.0824  | 0.0347  | 0.0572  | -0.0155 | -0.0076 | -0.2579 | 1.000   |        |       |
| GDP  | -0.2707 | -0.1185 | -0.1675 | -0.0433 | -0.1087 | 0.0353  | 0.0074  | 0.0795  | -0.2519 | 1.000  |       |
| EXH  | -0.1988 | -0.0656 | -0.1889 | -0.071  | 0.1234  | -0.0772 | 0.0240  | 0.6507  | -0.4259 | 0.4869 | 1.000 |

*Model Two: Profitability Proxy ROE*

The empirical results suggests that non-interest income, deposit size and inflation have positive association whereas operating expense, capital ratio, loan outstanding, market share, bank size and GDP growth has negative correlation with ROE under both POLS and RE model. POLS method finds operating expense to have inverse relation but RE model finds it positively associated. Market share identified to be significant under both methods, but non-interest income is significant just under RE model.

**Table 6: Model Two Estimation Results**

| Variable | Pooled Ordinary Least Square | Random Effect Model |
| --- | --- | --- |
| NII | 1.1540 (0.7708) | 1.8476** (0.651) |
| DPST | 0.1579 (0.1568) | 0.0298 (0.1297) |
| OPEX | -0.1156 (0.8323) | 0.4722 (0.7135) |
| CAPR | -0.2593 (0.1437) | -0.0276 (0.1298) |
| LTAR | -0.0434 (0.0245) | -0.0371 (0.0235) |
| MKT | -3.0845** (1.1819) | -1.6975* (0.6641) |
| SIZE | -0.0049 (0.0146) | -0.0033 (0.0159) |
| INF | 0.8462 (0.5149) | 0.7076 (0.5626) |
| GDP | -1.6159 (1.235) | -1.1628 (1.3375) |
| EXH | -0.1026 (0.1587) | -0.1739 (0.1728) |

**1% Significance
*5% Significance

**Table 7: Correlation Matrix of Explanatory Variables**

|      | ROE     | NII     | DPST    | OPEX    | CAPR    | LTAR    | MKT     | SIZE    | INF     | GDP    | EXH   |
|------|---------|---------|---------|---------|---------|---------|---------|---------|---------|--------|-------|
| ROE  | 1.000   |         |         |         |         |         |         |         |         |        |       |
| NII  | 0.1846  | 1.000   |         |         |         |         |         |         |         |        |       |
| DPST | 0.0250  | -0.1068 | 1.000   |         |         |         |         |         |         |        |       |
| OPEX | 0.0672  | 0.1775  | -0.0689 | 1.000   |         |         |         |         |         |        |       |
| CAPR | 0.0058  | 0.0783  | -0.2044 | 0.0401  | 1.000   |         |         |         |         |        |       |
| LTAR | -0.0611 | -0.0377 | 0.0660  | -0.1020 | -0.099  | 1.000   |         |         |         |        |       |
| MKT  | -0.1602 | -0.0283 | 0.0609  | 0.0712  | -0.0269 | -0.1160 | 1.000   |         |         |        |       |
| SIZE | -0.1669 | -0.0003 | -0.1464 | -0.0026 | 0.1372  | -0.1318 | 0.3773  | 1.000   |         |        |       |
| INF  | 0.1593  | 0.0820  | 0.0824  | 0.0347  | 0.0573  | -0.0156 | -0.0076 | -0.2578 | 1.000   |        |       |
| GDP  | -0.1582 | -0.1185 | -0.1674 | -0.0433 | -0.1088 | 0.0353  | 0.0074  | 0.0795  | -0.2519 | 1.000  |       |
| EXH  | -0.1922 | -0.066  | -0.1889 | -0.0709 | 0.1234  | -0.0771 | 0.0240  | 0.6507  | -0.4259 | 0.4869 | 1.000 |

*Model Three: Profitability Proxy NIM*

The POLS methods exhibit that non-interest income operating expense, market share and GDP growth has positive relationship with NIM whereas deposit size, capital ratio, loan outstanding, bank size and inflation has negative relation. The results from RE model suggest that non-interest income operating expense, capital, market share, loan outstanding and GDP growth is positively associated with profitability where deposit size, bank size and inflation have negative relation. POLS method finds that non-interest income, market share, and bank size is significant for profitability. In RE model non-interest income, market share, bank size and exchange rate proving to be significant.

**Table 8: Model Three Estimation Results**

| Variables | Pooled Ordinary Least Square | Random Effect Model |
| --- | --- | --- |
| NII | 0.9493** (0.0146) | 0.9803** (0.0124) |
| DPST | -0.0023 (0.0029) | -0.0019 (0.0025) |
| OPEX | 0.0173 (0.0158) | 0.0200 (0.0137) |
| CAPR | -0.0003 (0.0027) | 0.0007 (0.0025) |
| LTAR | -0.0002 (0.0005) | 0.0001 (0.0004) |
| MKT | 0.0796** (0.0224) | 0.0447** (0.0124) |
| SIZE | -0.0022** (0.0003) | -0.0022** (0.0003) |
| INF | -0.0008 (0.0098) | -0.0023 (0.0099) |
| GDP | 0.0103 (0.0234) | 0.0167 (0.0236) |
| EXH | 0.0188** (0.003) | 0.0188 (0.003) |

**1% Significance
*5% Significance

**Table 9: Correlation Matrix of Explanatory Variables**

|      | NIM     | NII     | DPST    | OPEX    | CAPR    | LTAR    | MKT     | SIZE    | INF     | GDP    | EXH   |
|------|---------|---------|---------|---------|---------|---------|---------|---------|---------|--------|-------|
| NIM  | 1.000   |         |         |         |         |         |         |         |         |        |       |
| NII  | 0.9689  | 1.000   |         |         |         |         |         |         |         |        |       |
| DPST | -0.1182 | -0.1068 | 1.000   | .       |         |         |         |         |         |        |       |
| OPEX | 0.1859  | 0.1775  | -0.0689 | 1.000   |         |         |         |         |         |        |       |
| CAPR | 0.0757  | 0.0783  | -0.2044 | 0.0402  | 1.000   |         |         |         |         |        |       |
| LTAR | -0.0338 | -0.0377 | 0.0660  | -0.1020 | -0.0990 | 1.000   |         |         |         |        |       |
| MKT  | -0.0309 | -0.0283 | 0.0609  | 0.0712  | -0.0269 | -0.1161 | 1.000   |         |         |        |       |
| SIZE | -0.0435 | -0.0003 | -0.1464 | -0.0026 | 0.1372  | -0.1318 | 0.3773  | 1.000   |         |        |       |
| INF  | 0.0563  | 0.0820  | 0.0824  | 0.0346  | 0.0573  | -0.0155 | -0.0076 | -0.2579 | 1.000   |        |       |
| GDP  | -0.0500 | -0.1185 | -0.1675 | -0.0434 | -0.1087 | 0.0353  | 0.0074  | 0.0795  | -0.2519 | 1.000  |       |
| EXH  | -0.0212 | -0.0659 | -0.1889 | -0.0709 | 0.1234  | -0.0772 | 0.0240  | 0.6507  | -0.4259 | 0.4869 | 1.000 |

*Discussions*

Empirical tests results exhibit that exogenous variables change sign based on the changes of endogenous variables. Although all the endogenous variables are proxy of profitability measurement, the calculation process and variances affected the relationship with explanatory variables.

Non-interest income is a supplementary source of earnings for the banks. Efficient allocation in resources in this segment can enhance earnings as well as profitability. Deposit is hypothesized to be positively correlated with profits, but the study revealed that not all the deposits are profitable for the banks. Deposits are associated with interest payment to the customers and lower investment opportunity can curtail bank's profit. The asset structure and investment diversification determine the earning trend and growth opportunities. The operating expense ratio shows the efficiency of the bank's management by keeping costs low while generating higher earnings. Operating expenses such as expansion of brank's branches, recruiting extra manpower for better services, aggressive sales and marketing etc. can drive the costs up while boosting earnings at the same time. Higher equity supports banks from the potential losses from risky investments. Higher capital and equity provide cushion against adverse financial conditions. Holding substantial funds can reduce the loanable funds and create liquidity excessiveness. The cost of equity is higher than the costs of other sources of financing which can drag the profit downwards. In the theoretical bank profit model composition of all outstanding loans and the survivor rate of these loans accounts for majority of profits. Higher outstanding loans and advances also enhance the chances of greater non-performing loans and provision requirement. Thus, overall balance between loan outstanding and non-performing loan ratio is a key element in profitability. Large bank size and higher market share provide access to higher liquidity, investments, and financing facilities. Larger banks and market leaders generally face complex new regulations and low growth opportunities in the industry which decreases marginal profit. Macroeconomic variables such as GDP, inflation and exchange rate can affect both supply and demand shock, thus affecting the profitability. The long run strategies and policies of the banks play the prime role determining the direction of profitability.

## IV. COVID-19 And BANKING PROFITABILITY

The outbreak of Covid-19 adversely affected both the banking industry and the economy. In 2020, demand and time deposits grew by 25.06% and 12.16%, respectively, compared to 2019. Loans and advances grew by just 9.05% in 2020 which is far less than projected. Weighted Average Rate of Interest on Deposits and Advances decreased by 89 basis points to 2.98% in December 2020 compared to the same period of the previous year. The plummeted demand for loans and advances shrinks the interest rate spread in the year 2020. The expanded gap between deposits and loans &

advances growth forced the banks to hold more liquidity and reduction in profitability. (Ahamed, 2021).

The disbursement of agricultural and non-farm rural credit decreased by 2.27% and 8.43%, respectively, during July'19-June'20 period compared to July'18-June'19. For the July-November'2020 period, agricultural credit increased by 11.28%, and non-farm rural credit decreased by 6.25% compared to the same period of the previous year. Disbursement and recovery of industrial term loans in the July-September'2020 period was 29.65% and 45.53%, respectively, compared to the same period of the preceding year. Cottage, Micro, Small & Medium Enterprises loan disbursement decreased by 27.60% in the July-September'2020 period equated to the prior period. (Bangladesh Bank, 2020).

The Covid-19 impacted the business conditions and negatively affected the repayment capacity of the borrowers. Bangladesh Bank relaxed the existing loan classification rules and formulated a special policy on loan rescheduling. The central bank took steps to defer the installments, suspension of interest charges, and subsidized interest expenses to the borrowers. Numerous refinancing schemes were also initiated to provide liquidity in the banking sector and support the borrowers. This enhanced amount of liquidity along with low credit demand shrinks the opportunity to generate enough return to cover the cost of capital.

## V. CONCLUSION

The study examined the relationship between bank profitability and a comprehensive list of bank-specific, industry-specific and macroeconomic variables using unique panel data from 23 Bangladeshi banks with large market shares from 2005 to 2019 employing the Pooled Ordinary Least Square (POLS) Method for regression estimation. The random effect model was used to check for robustness. Three variables (ROA, ROE, and NIM) were used as bank profitability proxies following standard literature. It was concluded that non-interest income, capital ratio, market share, bank size, GDP growth, and real exchange rates were significant to differing degrees in explaining the profitability of the sampled Bangladeshi banks.

The chief policy implication of this study is that in a regimented industry where price competition is not a viable source of competitive advantage, empirical test results exhibit that efficient allocation of resources to a supplementary source of earnings, non-interest income, can significantly enhance earnings as well as profitability. Banks that will pursue this income-source diversification strategy will be a gainer in the longer term and receive favorable ratings from investors and analysts. Large bank size and higher market share provide access to higher liquidity, investments, and financing facilities. Although this study found the mixed direction of the sign of the predictor variable, larger banks and market leaders generally face complex new regulations and low growth opportunities in the industry, which decreases marginal profit. This finding has important implications for managers running banking operations in Bangladesh as they navigate through the new regulations-growth nexus in a way that optimizes profitability. Business environment variables such as GDP growth and exchange rate can affect supply and demand shock, affecting profitability. Therefore, policymakers in charge of the country's macroeconomic management, especially the central bank, need to formulate economic growth-focused policies and manage the real exchange rate that works in favor of bank profitability.

Although the deposit is hypothesized to be positively correlated with profits, this study revealed that not all the deposits are profitable for the banks. A proxy for managerial efficiency, operating expenses such as the expansion of brank's branches, recruiting extra workforce for better services, aggressive sales and marketing, etc. can drive the costs up while boosting earnings at the same time. Higher capital and equity provide a cushion against adverse financial conditions. The cost of equity is higher than the costs of other sources of financing, which can drag the profit downwards. Higher outstanding loans and advances also enhance the chances of greater non-performing loans and provision requirements. Thus, the overall balance between loan outstanding and non-performing loan ratio is an element in profitability. Holding substantial funds can reduce the loanable funds and create liquidity excessiveness. However, interestingly, none of the three profitability proxies was affected by the liquidity variable in the current context. It is consistent with a previous study by Hossain and Ahamed (2015) that concluded that this phenomenon indicates Bangladeshi banks not pursuing systematic and modern Balance Sheet management strategies.

Finally, although this study provided a strong foundation in understanding the determinants of Bangladeshi banks' profitability, it has its limitations. There are some structural issues rather unique to the Bangladeshi banking industry like the six-nine interest rate regime fixed arbitrarily by the chief executive of the country, absence of natural functioning of the market forces in allowing poorly performing banks to go bankrupt or merge with banks with stronger fundamentals, exogenous influence on the bank management in disbursing low-quality assets etc. that need to be tackled separately to better understand the issue at hand. This is a crucial direction for future studies.

# CONFLICTS OF INTEREST STATEMENT

On behalf of all authors, the corresponding author states that there is no conflict of interest.